# A New Horizon of Data Communication through Quantum Entanglement

S.M. Rashadul Islam[1], Md. Manirul Islam[2] and Umme Salsabil[3]

[1] American International University-Bangladesh, Dhaka, Bangladesh
[2] American International University-Bangladesh, Dhaka, Bangladesh
[3] British Columbia Institute of Technology, Burnaby, BC, Canada
manirul@aiub.edu

**Abstract.** By the blessing of our existing data communication system, we can communicate or share our information with each other in every nook and corner of the world within some few seconds but there are some limitations in our traditional data communication system. Every day we are trying to overcome these limitations and improve our systems for better performance. Among them some problems may not be resolvable, for the reason of very basic or root dependencies of physics. In this paper, we have clarified some main drawbacks in our traditional communication system and provided a conceptual model to overcome these issues by using mystic Quantum Entanglement theorem rather than classical or modern physics phenomenon. In the end, we introduced a possible Quantum circuit diagram and Quantum network architecture for end-to-end data communication. It is predicted that through this hypothetical model data can be transmitted faster than light and it will be 100% real time between any distances without any kinds of traditional communication medium that are being used to date.

## 1 Introduction

The innovation of wheel is treated as one of the greatest achievements for human civilization because by reaching this milestone mankind first experienced that they can move from one place to another without using their feet. That was the beginning, now we can move faster than sound and trying to move at light speed. Our digital world is one step ahead because we can transmit our data at light speed using optical fiber medium. Now we are trying to transfer our data faster than light. But the question is, is it possible? If we want to send or transmit data faster than light, then we have to bypass all classical and modern physics laws. In traditional data communication system, we send data unit as a sequence of 0s or 1s. In between the sender and receiver, data travel through different medium using different type of devices, protocols and algorithms obligations. In this long journey, data can be mismatched or lost and even sometimes cannot reach its desired destination. Sometimes it is observed that one device’s performance or throughput can hamper the total communication system. The traditional communication system is not flexible as it depends on the following key factors:

- Data transmission time
- Data transmission rate
- Data integrity
- Data transmission medium
- Number of Algorithms, protocols, applications etc.
- Number of intermediate devices.

## 2 Literature Review

Quantum Mechanics is a concept which first introduce in physics around 100 years ago, which does not follow classical physics laws. Quantum theory is the theoretical basis of modern physics that explains the nature and behavior of matter and energy on the atomic and subatomic level. The nature and behavior of matter and energy at that level is sometimes referred to as quantum physics and quantum mechanics. By quantum physics we can mainly describe quantum particles behaviors. Quantum particles are mainly two types Fermions and Bosons. Fermions is responsible for electron, proton and neutron creation and Bosons is responsible for creating Protons, W bosons, Z boson, Gluon and Higgs Boson [3].

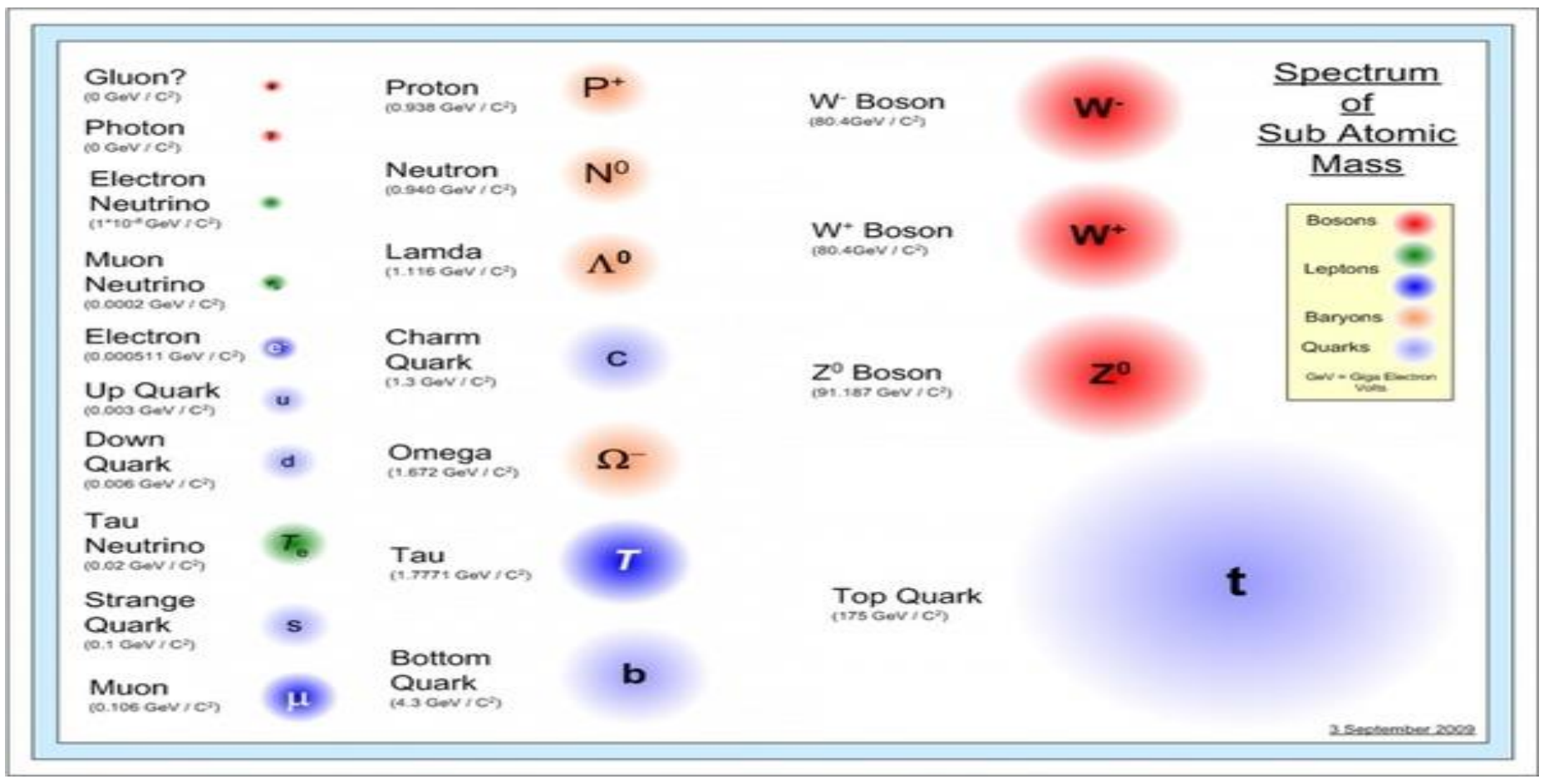

**Fig. 1.** Spectrum of Sub Atomic Mass

All fundamental particles have a property called Spin. According to quantum physics, this particle doesn’t have a well-defined spin at all. Particles are normally in random spin process. Particles are phasey and undefined, until they observed, and they are always entangled with each other. If the particles are measured in the same direction, then their spin must be opposite [4]. In quantum physics, the Quantum entan-glement of particles describes a relationship between their fundamental properties that is not possible coincidentally. This could refer to states such as their momentum, position, or polarization [5]. Knowing something about one of these characteristics for one particle will describe something about the same characteristic for the other.

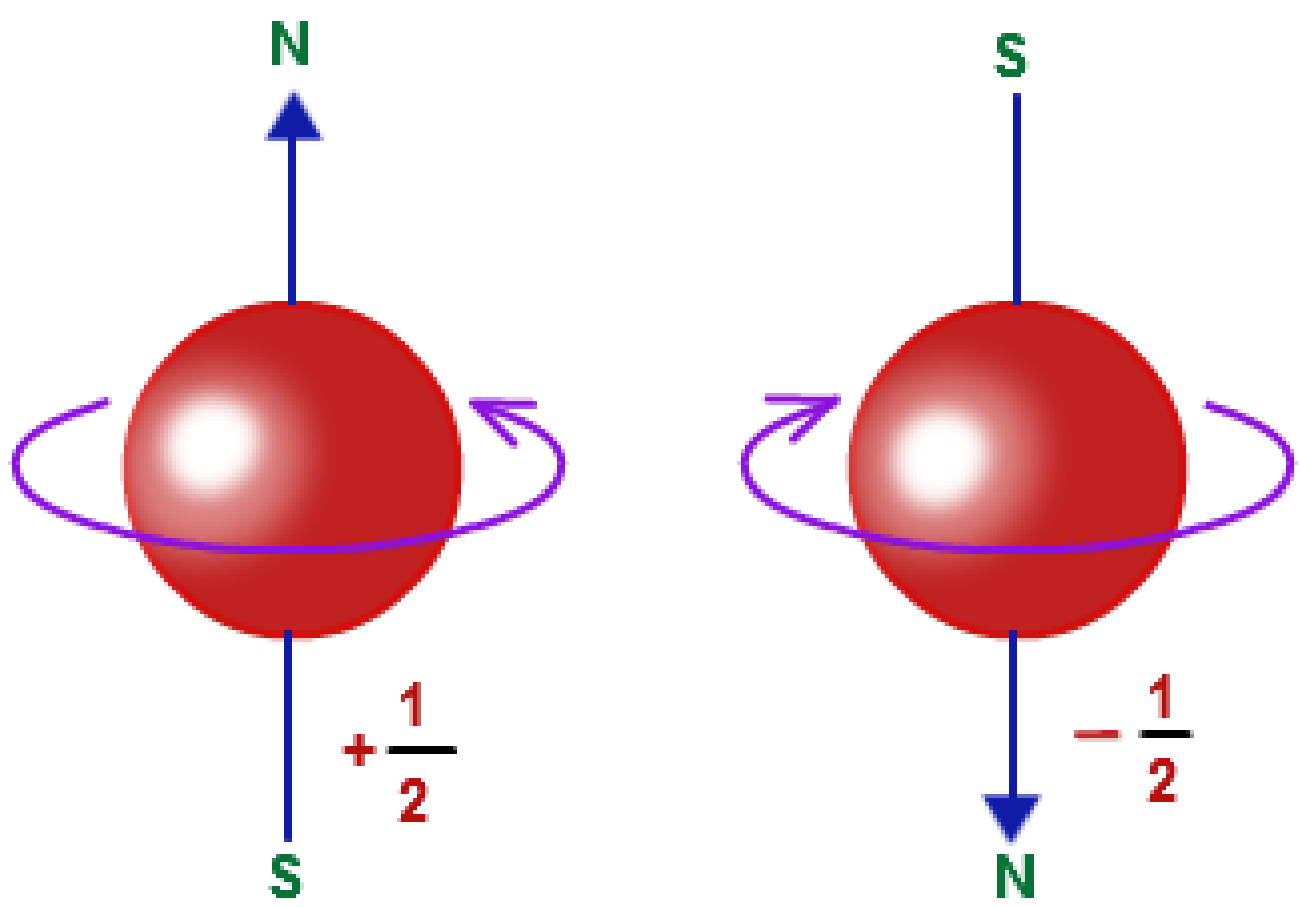

**Fig. 2.** Entangled Quantum Particle Spin Axis

Quantum Entanglement is dubbed “Spooky action at a distance” by Albrert Einstein, described a situation where particles can remain connected such that the physical properties of one will affect the other, no matter the distance (thousand lightyears) between them [6]. Einstein did not like the idea, since it was violating classical descriptions of the world. So, he proposed one way that entanglement could coexist with classical physics, if there exists an unknown, "hidden" variable that acted as a messenger between the pair of entangled particles, keeping their fates entwined. There was no way to test whether Einstein's view or the stranger alternative, in which particles "communicate" faster than the speed of light and particles have no objective state until they are observed was true. Finally, in the 1960s, physicist Sir John Bell came up with a test that disproves the existence of these hidden variables which would mean that the quantum world is extremely weird. Recently, a group at the University of Glasgow used a sophisticated system of lasers and crystals to capture the first-ever photo of quantum entanglement violating one of what's now known as "Bell's inequalities." This is "the pivotal test of quantum entanglement" said senior author Miles Padgett, though people have been using quantum entanglement and Bell's inequalities in applications such as quantum computing and cryptography [2]. To take the photo, Padgett and his team first had to entangle photons, or light particles, using a tried-and-true method. They hit a crystal with an ultraviolet (UV) laser, and some of those photons from the laser broke apart into two photons. "Due to conservation of both energy and momentum, each resulting pair photon are entangled," Padgett said.

They found that the entangled pairs were correlated, or in sync, far more frequently than it was expected if a hidden variable were involved. In other words, this pair violated Bell's inequalities. The researchers snapped a picture using a special camera that could detect individual photons, but only took a photo when a photon arrived with its entangled partner [1].

## 3 Research Goals

Particles are normally in random spin process. They are phasey and undefined until they are observed. When observed, both of them have the same properties. If we observe two entangled quantum particle's spin properties at the same time, they are always at the opposite direction with each other. In this paper, we are going to use Entangled quantum particle's physical properties "SPIN" law for data transmission.

If we observe an entangled particle's SPIN position and presume that position's value is '0', then we know that other entangled Quantum particle's SPIN direction is opposite, let's give this spin position value "1". By reversing the value "1", we get "0". In this process, we don't need any medium and that is 100% real time. In that way, we can solve all kinds of Time Complexity, Distance Complexity and medium complexity for data Transmission System.

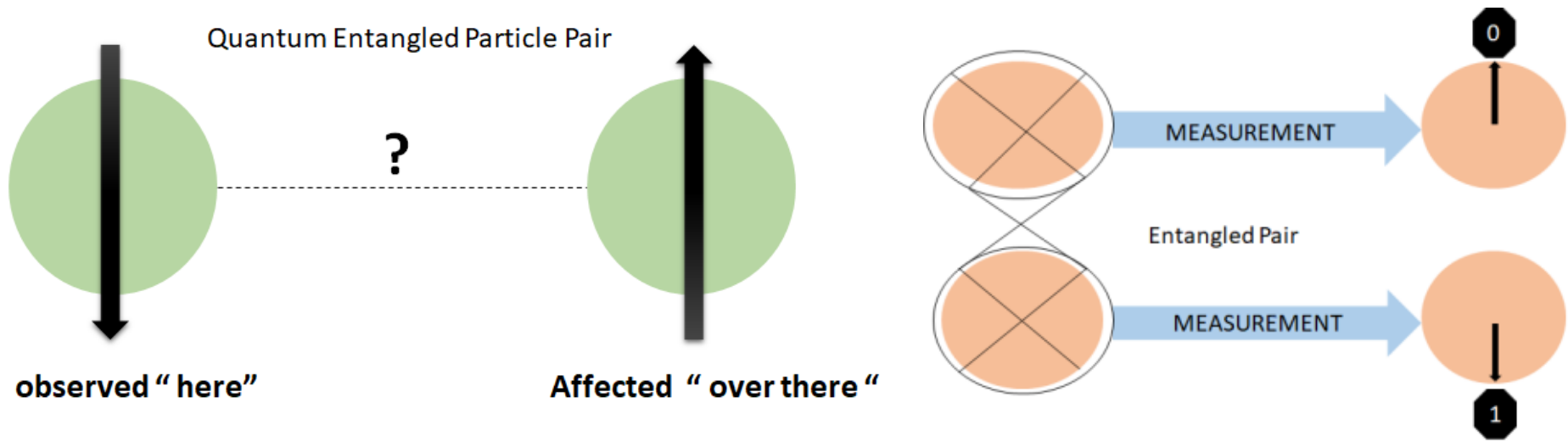

**Fig. 3.** Observation of Quantum Particles Spin

Based on the above principles and acknowledging the quantum advancements, the main research objectives and goals are given below.

- Methodologies to test quantum computation enabled devices and transmit their data to the distinct devices [8].
- New algorithms, protocols and applications for quantum networks [7].
- Overcome distance complexity for data communication.
- Overcome medium complexity for data transmission.
- Overcome time Complexity for data transmission.
- Realtime Communication without any latency and jitter.
- Overcome Bandwidth complexity for data transmission.
- Increase data security and integrity.

# 4 Design Architecture of the Hypothetical Model

To explain the hypothetical model, it is necessary to describe how sender and receiver transmit data obeying Quantum Entanglement phenomena. Both sender and receiver contain two individual circuits i.e. Tx Plate and Tx Circuit, Rx Plate and Rx Circuit.

## 4.1 Sender Tx Plate and Circuit

In a sender node, there are quantum entangled particle's Tx plate which communicates with a compiler and compiler interacts with quantum processor. Quantum processor process the data and send it back to Tx compiler. Compiler receives the data and send corresponding signal to ionized quantum particle in Tx Plate.

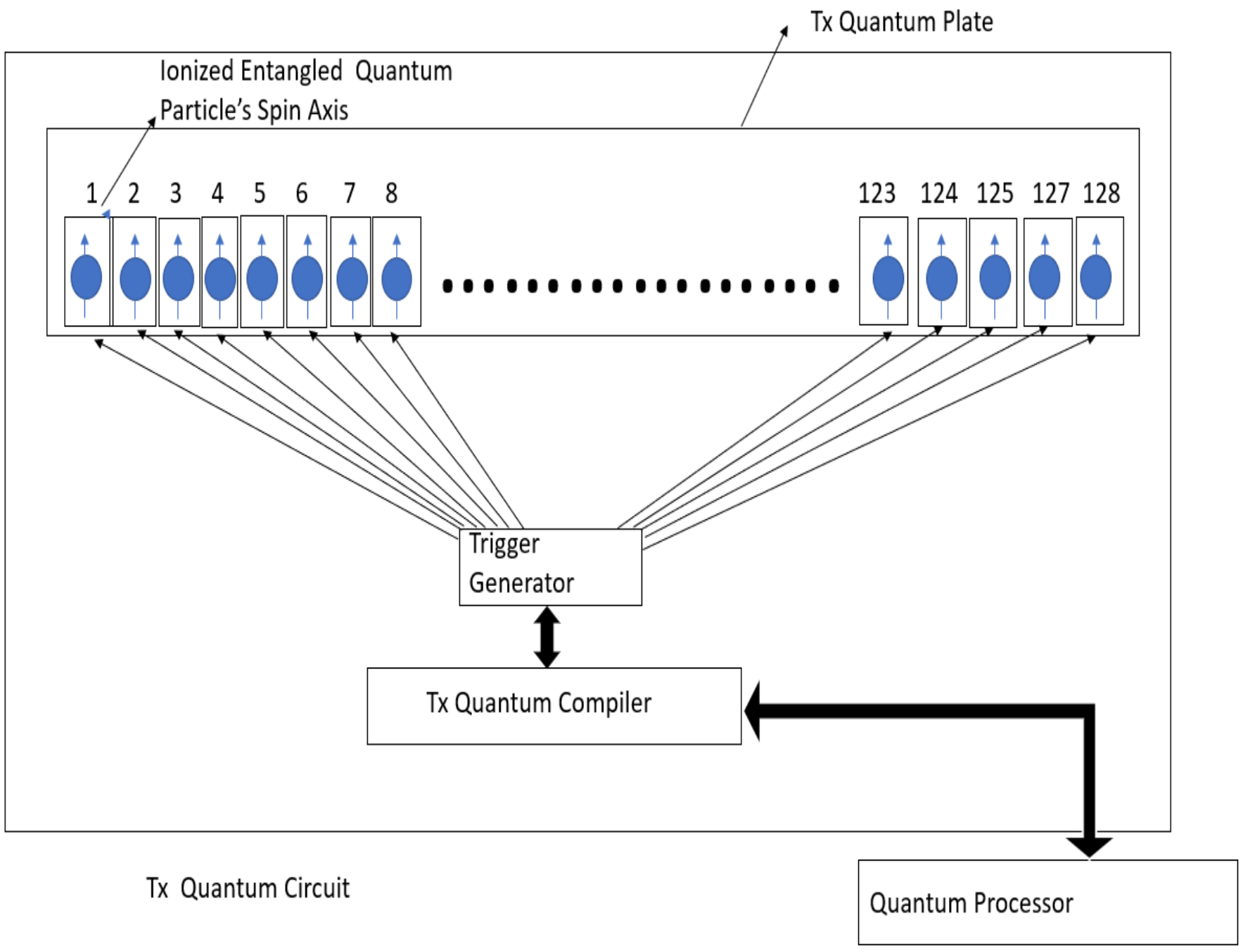

**Fig. 4.** Tx Circuit Diagram of Entangled Ionized Quantum Particle at Sender Side

1. Every plate has 128 ionized quantum entangled particles.
2. Ionization gives fix axis direction of the Quantum entangled particle's spin properties.
3. When Rx compiler wants to send a value to the plate, it produces corresponding signal to the ionized particle to trigger the spin direction. It will generate an opposite Spin at the entangled particle on the other side.

### 4.2 Receiver Rx Plate and Circuit

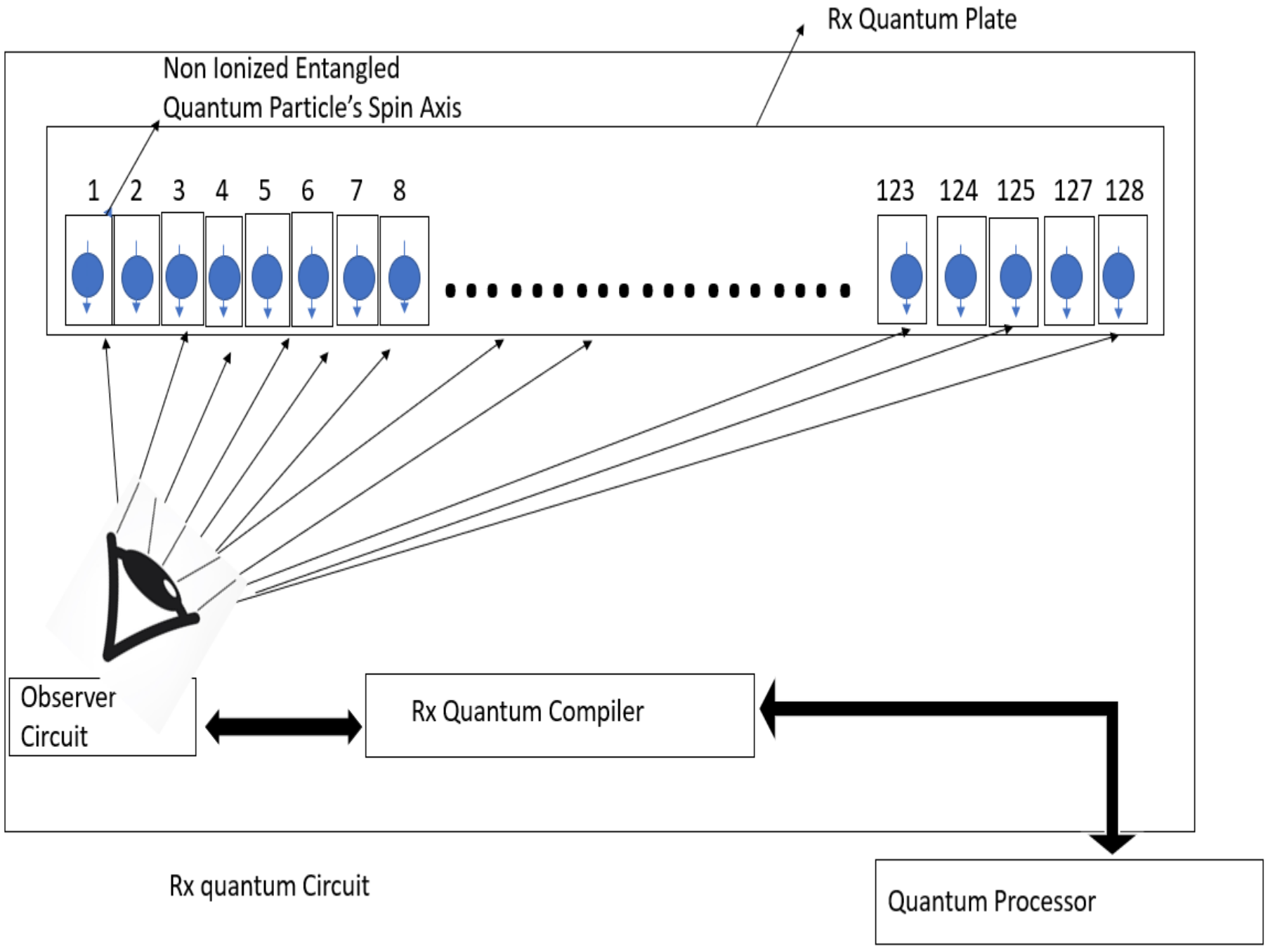

**Fig. 5.** Rx Circuit Diagram of Entangled Ionized Quantum Particle at Receiver Side

1. In the same way receiver side Rx plate contain 128 non-ionized entangled quantum particles which spin axis is always opposite direction of sender Tx particles Spin axis (when it is observed).
2. Every Rx circuit have an observer to observe particles SPIN axis direction and continuously send observation result to the compiler.
3. Compiler analyses the result and decode the received data. After that compiler sends the data to receiver processor.

A full diagram of sender and receiver side will give a clearer view of the communication process.

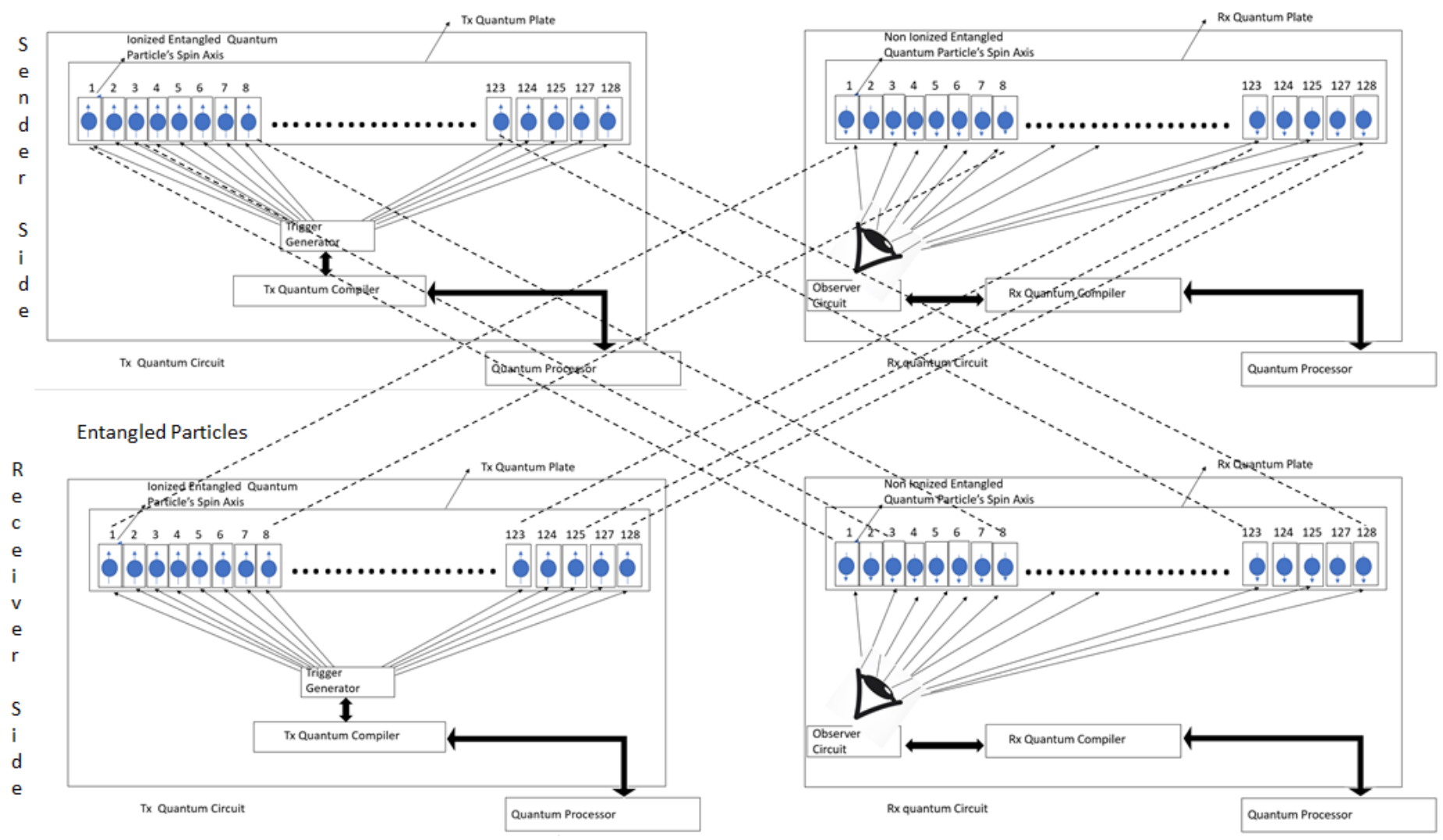

**Fig. 6.** Full Circuit Diagram of a Communication Circuit

## 5 Proposed Quantum Network Diagram

In this section, we have proposed Quantum Network Diagram for the World and the Universe.

### 5.1 Quantum Network Diagram for the World

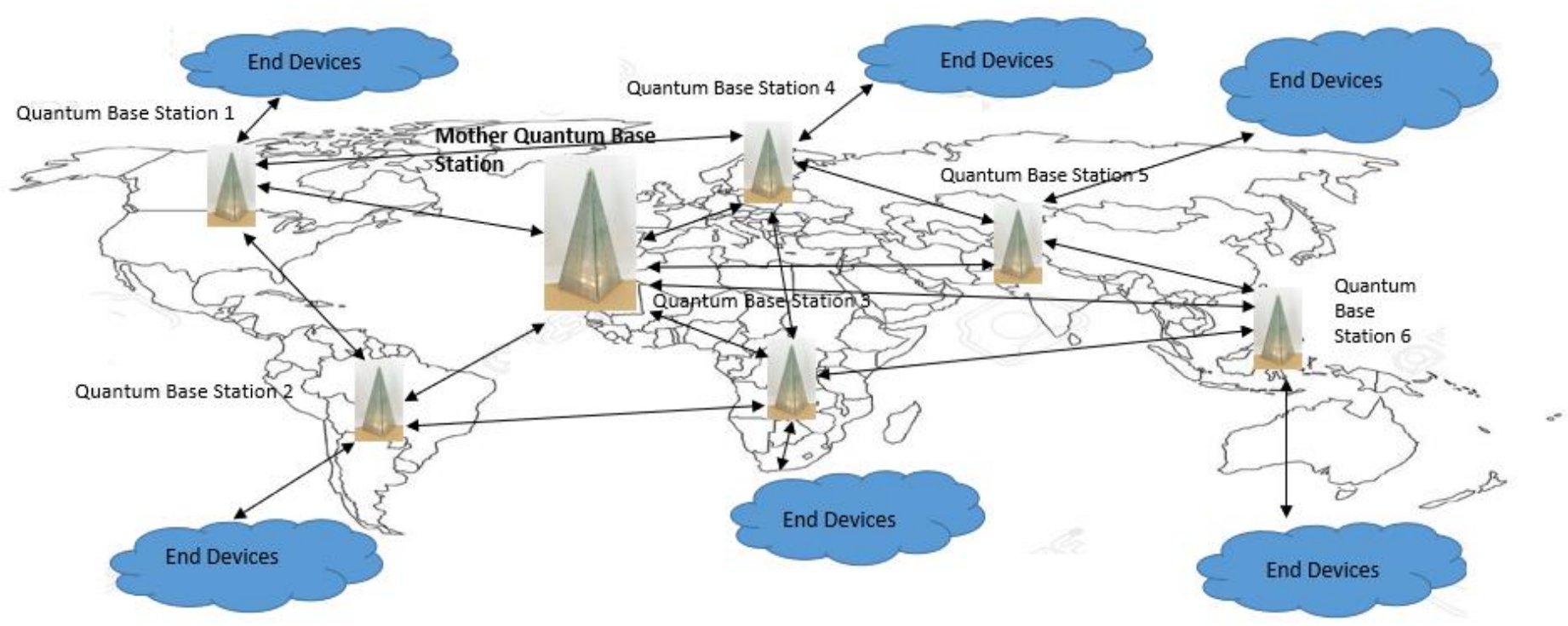

**Fig. 7.** Quantum Network Diagram for the World

There are two types of Quantum Base Station: (i) Mother Quantum Base Station (ii) Child Quantum Base Station.

- Each end device is individually paired with a specific Child Quantum Base Station.
- Every Child Base Station is responsible to provide desired information and services for the end devices.
- Every Child Base Station is connected with Mother Quantum Base Station.

### 5.2 Quantum Network Diagram for the Universe

Soon mankind will set their colony in Mars and someday will set their feet on the other side of the galaxies. Inter-Planet and Inter-Galactic communications will be required by then. Every planet will have a Mother Quantum Base Station and this station will maintain communication with other planet's Mother Quantum Base station.

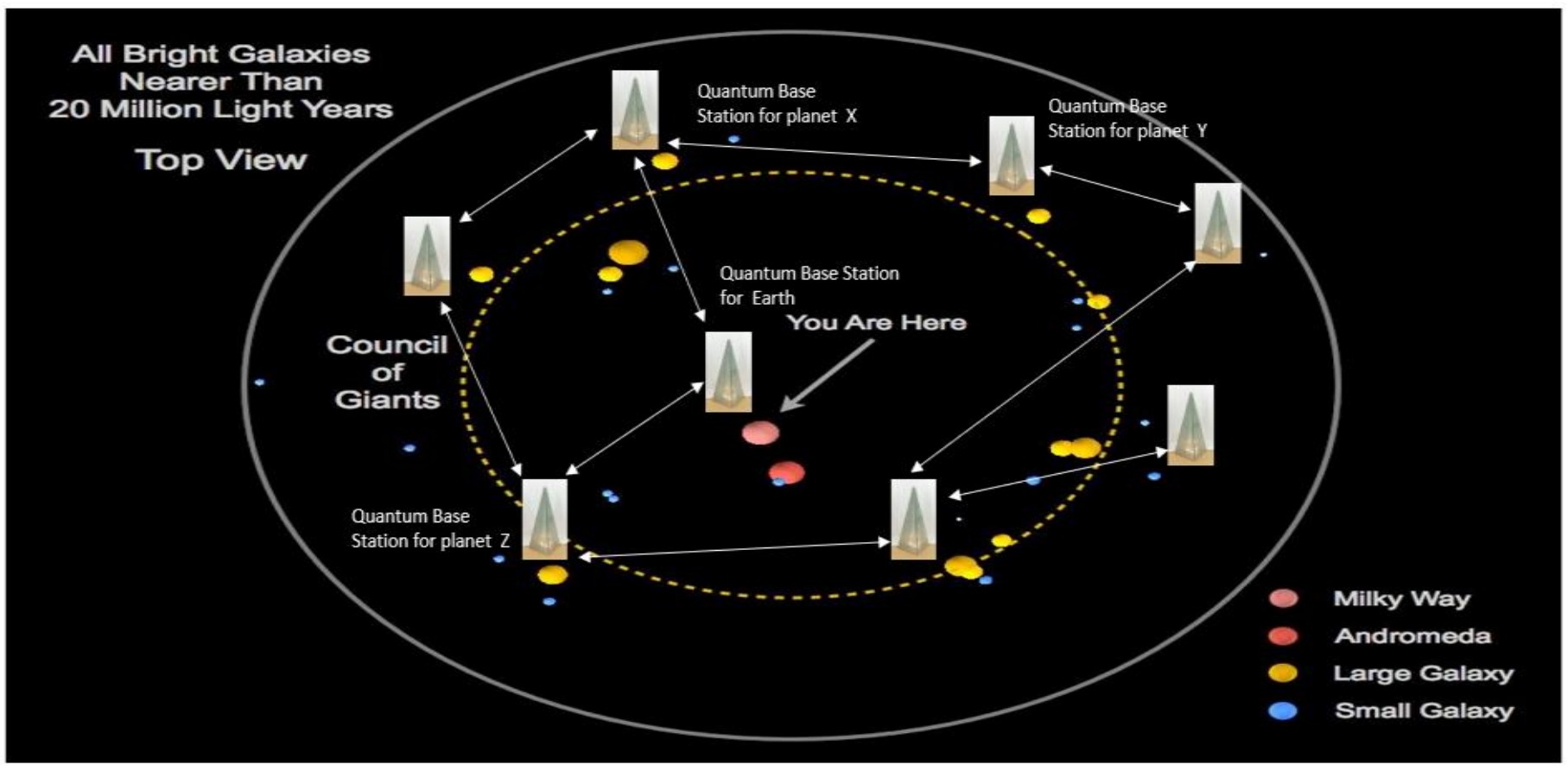

**Fig. 8.** Quantum Network Diagram for the Universe

## 6 End-to-End Data Flow Process for Quantum Network

In this section, we will describe two scenarios on how communication will be established between end-devices under same Quantum Base Station and between different Quantum Base Station. Every end user will have a unique Quantum Circuit ID (QID) and they will be directly connected with a Quantum Base station with the help of corresponding Entangled Quantum Circuit. This circuit will be responsible for data communication between end-users to stations and stations to end-users.

### 6.1 End-user to End-user Communication Process within Same Quantum Base Stations

(i) Let, User-A and User-B having different QID (QID-A and QID-B) wants to transmit data and both of them are connected to the same Quantum Base Station (QBS-1).

(ii) User-A will send a request to QBS-1 to establish a session with User-B. After receiving the request, QBS-1 will search its own Database to locate the QID of User-B.
(iii) If QBS-1 finds the requested QID in its own Database, then QBS-1 will negotiate with User-B directly.
(iv) If User-B agrees, then QBS-1 will establish a data transmission session between the two.
(v) After completion of data transmission, QBS-1 will tear down the session between User-A and User-B.

### 6.2 End-user to End-user Communication Process between Different Quantum Base Stations

(i) Let, User-A and User-C having different QID (QID-A and QID-C) wants to transmit data between them. User-A is connected to Quantum Base Station 1 (QBS-1) and User-B is connected to Quantum Base Station 2 (QBS-2). Both QBS-1 and QBS-2 are entangled with a Mother Quantum Base Station.
(ii) User-A will send a request to QBS-1 to establish a session with User-C. After receiving the request, QBS-1 will search its own Database to locate the QID of User-C.
(iii) In this case, QBS-1 is unable to find the QID of User-C. So QBS-1 will now query the Mother Quantum Base Station to locate the QID of User-C. Mother Base Station is the central repository of all the QIDs of the planet. After locating the record in the database, Mother Base Station will inform QBS-1 that User-C is connected to QBS-2 through Quantum Entangled Circuit.
(iv) QBS-1 will now directly communicate with QBS-2 and express the desire that User-A wants to establish a session with User-C.
(v) QBS-2 will communicate with User-C and if User-C agrees, then data transmission will start. Communication flow will be User-A to QBS-1, QBS-1 to QBS-2, then QBS-2 to User-C and vice versa.
(vi) After finishing data transmission between, they will tear down the session and release the Quantum Circuit.

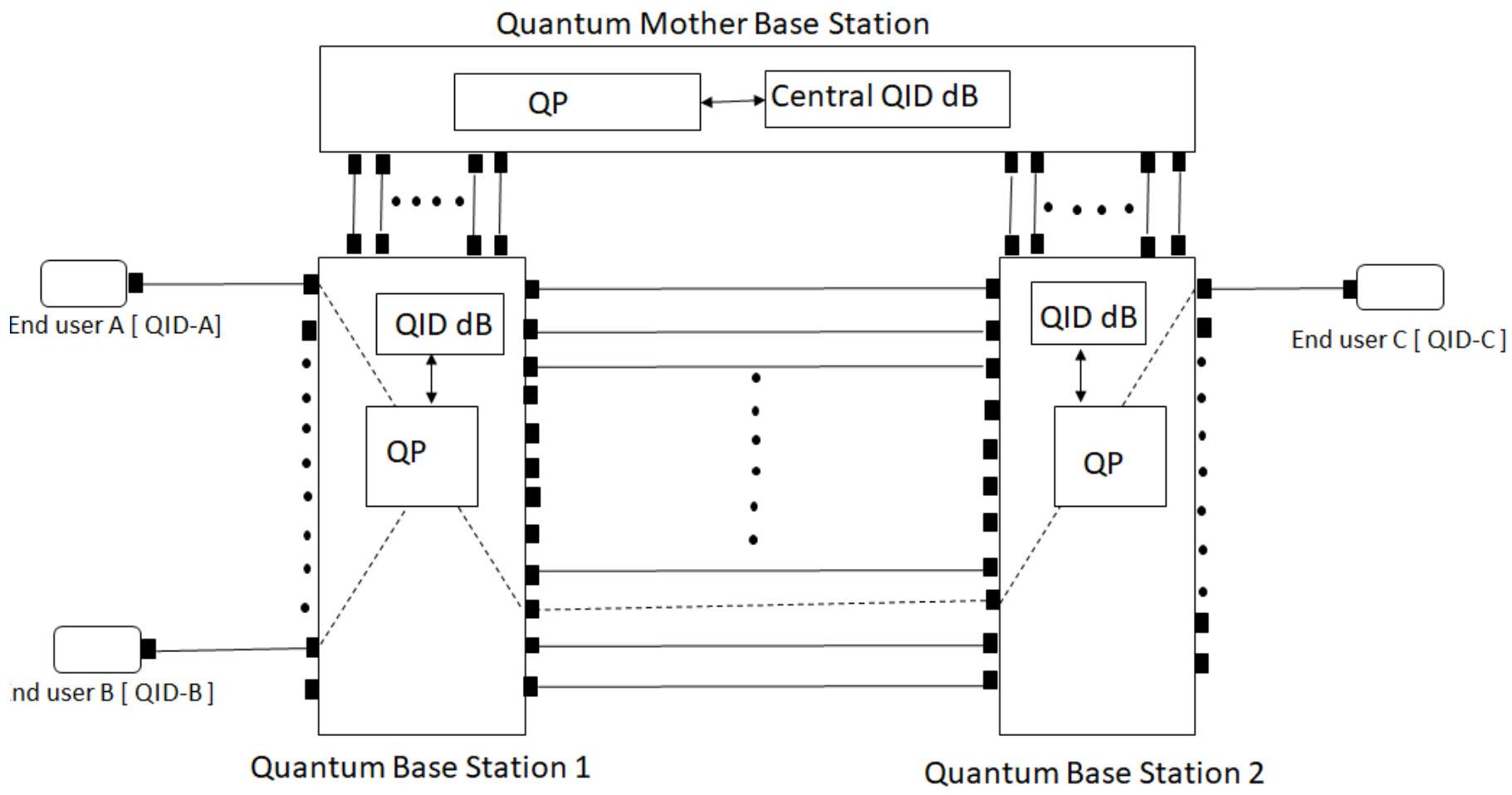

**Fig. 9.** End-to-End Communication Process

## 7 Challenges

There are several challenges identified:

- Trigger generation technique for Quantum particles' SPIN control
- Hardware limitation (modular or chipset or circuit)
- Algorithms/protocols for dispersedly-connected system.

## 8 Conclusion

The advancement of technology is meant for the betterment of mankind. It is very important for human civilization to explore beyond the world for the survival of its own race and future generations. For that, they must build new communication system which don't have any time, distance or medium complexity. In this study, we shed light on a conceptual model for Quantum communication enabled devices and a proposed network architecture for the whole universe. It is anticipated that this research will open up a new door for Quantum Entanglement and add a new dimension in this arena of research. Though, a tremendous amount of works has already been done to establish a fully functional quantum communication system in both physical and logical level, there are a lot of open questions and research challenges which are still in mystery. To reveal more about it, more study and research is needed on this domain.

## References

1. LiveScience, https://www.livescience.com/65969-quantum-entanglement-photo.html, last accessed 2019/12/25.
2. Kozlowski, W., Stephanie, W.: Towards Large Scale Quantum Networks. NANOCOM ’19. September 2019. https://arxiv.org/pdf/1909.08396.pdf, last accessed 2019/12/07.
3. whatIs.com, https://whatis.techtarget.com/definition/quantum-theory, last accessed 2019/12/05.
4. Veritasium, Quantum Entanglement & Spooky Action at a Distance. https://www.youtube.com/watch?v=ZuvK-od647c, last accessed 2019/12/05.
5. Science alert, https://www.sciencealert.com/entanglement, last accessed 2019/12/10
6. Public Radio International, https://www.pri.org/stories/2017-07-25/love-quantum-physics-and-entanglement, last accessed 2019/12/15.
7. CERN, https://cds.cern.ch/record/705724/files/0401076.pdf, last accessed 2019/12/21.
8. Towards Data Science, https://towardsdatascience.com/quantum-computing-and-ai-789fc9c28c5b, last accessed 2019/12/19.